\documentclass[preprint,12pt,authoryear]{elsarticle}

\usepackage{framed,multirow}
\usepackage{amssymb}
\usepackage{latexsym}
\usepackage{amsmath}
\usepackage{subcaption}
\usepackage{graphicx}
\usepackage{float}
\usepackage{url}
\usepackage{xcolor}

\makeatletter
\def\ps@pprintTitle{%
  \let\@oddhead\@empty
  \let\@evenhead\@empty
  \def\@oddfoot{\hfill\thepage\hfill}%
  \let\@evenfoot\@oddfoot}
\makeatother

\begin{document}

\begin{frontmatter}

\title{Towards Myocardial Infarction Diagnosis: An Investigation into Machine Learning and Deep Learning Algorithms using ECG Data}

\author{Rafid Ahnaf}
\ead{rafid.ahnaf@g.bracu.ac.bd}

\author{Husne Mubarak}
\ead{husne.mubarak@g.bracu.ac.bd}

\author{Fahim Ahsan}
\ead{fahim.ahsan@g.bracu.ac.bd}

\author{G. M. Refatul Islam}
\ead{gm.refatul.islam@g.bracu.ac.bd}

\author{Rafeed Rahman}
\ead{rahman.rafeed@bracu.ac.bd}

\author{Nabuat Zaman Nahim}
\ead{nabuat.zaman@bracu.ac.bd}

\address{Department of Computer Science and Engineering, BRAC University, Dhaka, Bangladesh}

\begin{abstract}
Myocardial Infarction (MI) is a severe and potentially fatal cardiac condition caused by myocardial injury due to ischemia, and timely detection can substantially reduce its mortality. Electrocardiography (ECG) is the primary diagnostic modality for MI, but manual interpretation is susceptible to human error. In this work, we propose Visionary V2, a multi-step, adaptive filtration method for ECG signal normalization, and use it to study machine learning (ML) and deep learning (DL) approaches to MI detection on the PTB Diagnostic ECG Database using all 15 available leads. On statistical features extracted from the filtered signals, such as the R-R interval, a Bagging ensemble of decision trees reaches 96.6\% accuracy, compared with 87.7\% when the signals are processed with BioSPPy. We further generate an image dataset from the filtered signals, on which a custom ResNet-50 misclassifies only 1 of 7,000 validation images and a custom CNN reaches an F1 score of about 98.7\%. Finally, we apply LIME, Grad-CAM and Integrated Gradients to verify that the image models attend to the ECG waveform rather than the background. These results indicate that careful signal normalization combined with ML and DL models can support fast and accurate automated MI screening.
\end{abstract}


\end{frontmatter}

\section{Introduction}
Myocardial Infarction (MI), commonly called ``heart attack,'' is a life-threatening cardiovascular condition caused by the complete or partial obstruction of the coronary arteries responsible for providing oxygenated blood to the cardiac muscle. MI affected heart muscles may introduce increased mechanical stress to the heart during muscle contraction for pumping blood which may result in structural and conformational changes like left ventricular remodeling. While symptoms like unconsciousness, breathing problems, and chest pain are common, many victims fail to exhibit the symptoms at an early stage. Additionally, factors such as biophysical conditions, previous medical conditions, and lifestyle choices can contribute to the development of MI.  

Given the high mortality rate associated with it, early detection is crucial for providing timely treatment. Traditional electrocardiogram (ECG) analysis, while effective, has limitations such as being subject to human error and misdiagnosis. It also can't identify all possible MI risk factors, leading to false negatives or positives. Research suggests that in cases of individuals presenting with significant symptoms of cardiac disease, cancer, and infection, 10\% are misdiagnosed or are already at an advanced stage of the disease. Furthermore, 53.9\% of these individuals have either become disabled or passed away as a result of medical errors or incorrect diagnosis. As a result, automated detection consisting of high-class accuracy and precision is a prerequisite for assisting medical practitioners with quick and accurate diagnosis of MI. This motivated us to explore the domain of how we can detect Myocardial Infarction in people of all genders and ages with more accuracy and efficiency within a short time than the traditional methods by utilizing modern machine learning and deep learning approaches. In our work, we came across several machine and deep learning models that can accurately classify ECG signals as indicative of MI or not both from numerical and image data derived from ECG. In order to denoise the raw ECG signal for better extraction of features, we also developed a dynamic robust pre-processing filter with the ability of adaptive filtering. Moreover, implementing explainable AI techniques in our work has brought the explication of the applied models to a higher dimension using our derived results. We believe this combination of accurate detection and interpretable outputs can meaningfully support clinicians in diagnosing MI.

\section{Background and Context}

Electrocardiograms (ECGs) provide a graphical representation of the electrical activity of the heart, providing detailed information about a person's cardiac health. A single ECG wave typically consists of several components—P wave, Q wave, T wave, QRS complex, and ST segment—that are critical for assessing cardiovascular abnormalities such as Myocardial Infarction (MI). To facilitate both medical evaluation and research, ECG data must be digitized using various methods aimed at automating the complex diagnostic process.

Deep learning (DL) and machine learning (ML) have emerged as potent tools for automated MI detection. For effective classification, conventional ML algorithms frequently require carefully preprocessed ECG waveforms and engineered features. To improve ML performance, researchers used advanced techniques such as wavelet transformation, principal component analysis, and linear discriminant analysis. These feature extraction methods, however, are labor-intensive and time-consuming. Deep learning models, particularly CNN, on the other hand, show promise in automatically extracting features from raw ECG data, simplifying the analytical workflow.

Explainable Artificial Intelligence (XAI) also plays a pivotal role in interpreting the complex models, providing a rationale for predictions. This is crucial in medical settings, where understanding the reasoning behind model’s decision for a diagnosis can have life-altering implications. XAI techniques applied to ECG data aim to provide interpretability and accountability in automated MI detection models, enhancing clinicians' trust in these systems.

Furthermore, medical professionals traditionally rely on manual evaluation of ECG, focusing on patterns like ST-segment elevation to identify cardiac abnormalities. However, manual methods are subject to error, supporting the need for automated systems that can achieve high diagnostic accuracy reliably.

\section{Literature and Technologies}

The accuracy and prompt diagnosis of Myocardial Infarction (MI) is of utmost importance for the patient's care and results. In recent times, there has been an increasing focus on the utilization of Machine and Deep Learning techniques to improve MI detection using ECG data. According to \cite{1}, manual and inaccurate detection of cardiac abnormalities may result in the loss of life of MI patients. For instance, expert cardiologists may overlook ST-segment elevations in as many as 18\% of cases, as these elevations can be subtle and atypical in certain patients, and difficult to differentiate from other EKG abnormalities. ML in Myocardial Infarction detection offers a number of benefits over manual methods, including consistency, speed, efficiency, 24/7 availability, etc., making it a valuable asset in healthcare.

A paper on Classification of Myocardial Infarction by \cite{2} has demonstrated the usefulness of CNNs in extracting features from ECG signals with multiple leads. This approach is based on the utilization of spatial information across several leads which enhances diagnostic accuracy. However, the study does not fully investigate the temporal dependencies of ECG data. By including temporal dependencies, it is possible to detect subtle variations in ECG patterns, increase sensitivity to early and atypical indicators of MI, as well as enhancing the overall precision of the diagnostic models. Inadequate consideration of temporal dependencies in MI detection can result in inaccurate diagnoses. A similar work on Detecting MI using ML methods by \cite{3} provides insights into the application of traditional machine learning algorithms in ECG analysis. The results of the study suggest that ML can be used to detect MI with high accuracy using only a single lead of the ECG but that alone dismisses chances of detecting other types of MI that are detected using other leads which results in misidentification. 

Another research by \cite{4} introduces the idea of multispectral concatenation to improve feature representation in MI detection using ECG data. The multispectral approach enhances MI detection by capturing a broader range of ECG data for more detailed analysis, thereby improving the model's accuracy in identifying fine signal variations. The paper highlights the critical role of feature engineering, preprocessing, and data representation in optimizing the pipeline. It also introduces an innovative method for ECG-based MI detection, offering valuable insights into the field. However, there are drawbacks to the proposed method as it may need to be evaluated across diverse datasets in order to provide a more robust and generalizable approach. The interpretability of a model’s decisions and their clinical implications is not discussed in detail, which may be a key consideration in real world healthcare applications. The limitations found in these researches on AI-based MI detection are quite visible and could have detrimental effects.

\section{Data Ecosystem}

\subsection{PTBDB Introduction}

In our research on Myocardial Infarction (MI) detection via ECG, we primarily utilized data from the PTBDB dataset \citep{10}. The main contributor to this dataset was Physikalisch-Technische Bundesanstalt (PTB), the National Metrology Institute of Germany. This data has been publicly available for research and benchmarking for complex physiological signals.

The PTBDB dataset comprises 549 ECG records from both male and female subjects aged between 17 and 87, with an average age of 61.6 years. These records were collected using a custom personal tape recorder with 16 channels, 14 dedicated to ECG, one for respiration, and one for line voltage. Each ECG record features signals from 15 leads, including 12 conventional leads (i, ii, iii, avr, avl, avf, v1, v2, v3, v4, v5, v6) and 3 Frank leads (vx, vy, vz). The signals have a sampling rate of 1000 ticks per second per lead as shown in Fig. 1.

\begin{figure}[htbp]
\centerline{\includegraphics[trim={0cm 25cm 0cm 0cm},clip,width=0.5\textwidth]{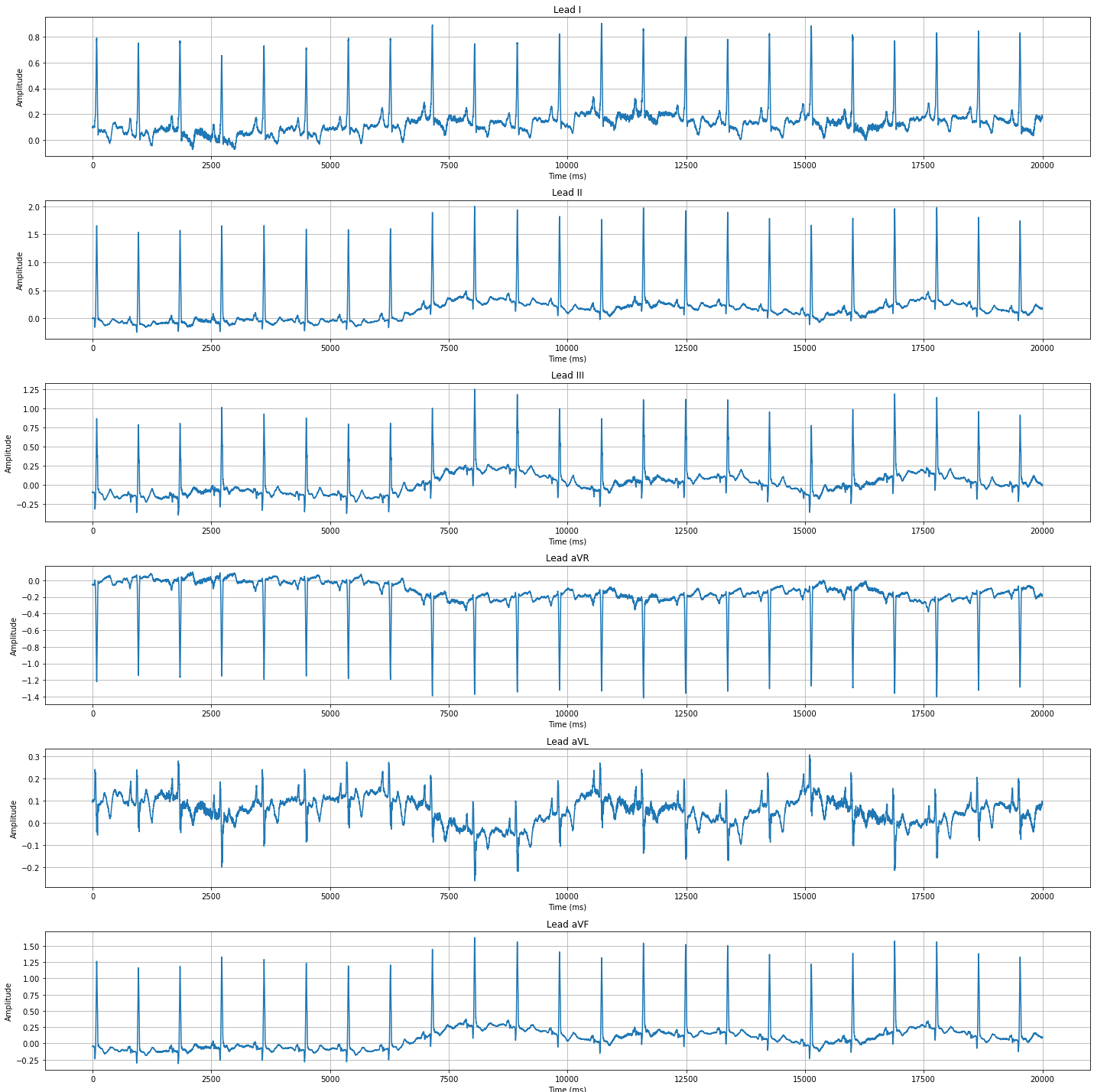}}
\caption{Visualization of unfiltered single lead readings}
\label{fig}
\end{figure}

The subjects were classified into multiple groups having different heart conditions from which we filtered out the patients with Myocardial Infarction and Healthy conditions only. In addition to ECG signal, the dataset provided detailed clinical summaries for each patient, including age, sex, diagnosis, medical history, medication, and relevant medical conditions.

\subsection{Preprocessing Comparison}

The initial step in processing biophysical signals like ECG is noise reduction, a complex task that also has the risks of important data loss. For comparing with our own filtration process, we initially implemented BioSPPy \citep{11}, a bio-signal processing toolbox, for noise reduction (Fig. 2). Once the noise reduction is applied, statistical factors like mean, standard deviation, minimum, maximum, skew and kurtosis reading of the ECG along with R-R intervals has been extracted as key features for MI detection. The conventional markers for detecting MI, i.e., ST segment elevation, QRS complex, P wave, T wave, cannot be extracted here due to the lack of any specific pointers in the signal. Since R-peaks are the extreme data points in the record they can be easily identified. Additionally, the ECG Heartbeat Categorization (EHC) dataset, widely available as open-source data for ECG testing, has also been used as a benchmark for model performance \citep{6}. The dataset is derived from Lead-2 ECG signals only. It is resampled at 125Hz, and focuses on a 10-second window featuring normalized amplitudes and calculated R-R intervals. It includes 290 subjects, of whom 148 have MI and 52 are healthy controls, while the rest have other cardiac conditions. Unlike EHC, our approach employs all 15 leads for feature extraction while encompassing a broader range of themes and health concerns.

\begin{figure}[htbp]
\centerline{\includegraphics[scale=0.5]{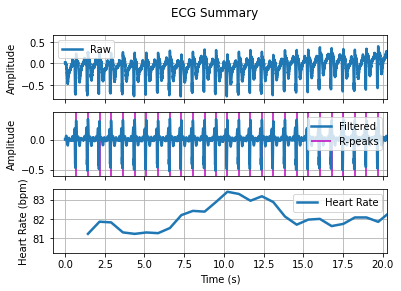}}
\caption{Raw and Filtered ECG with marked R-peaks using Biosppy}
\label{fig}
\end{figure}

\vspace{-8mm}

\section{Our Approach}

\subsection{Visionary V2}

\begin{figure*}[htbp]
    \centering
    \begin{subfigure}[b]{0.33\textwidth}
        \includegraphics[width=\textwidth]{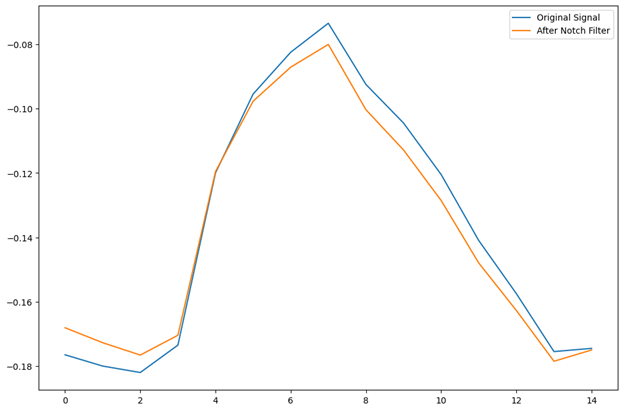}
        \caption{First Layer}
        \label{fig:image1}
    \end{subfigure}%
    \begin{subfigure}[b]{0.33\textwidth}
        \includegraphics[width=\textwidth]{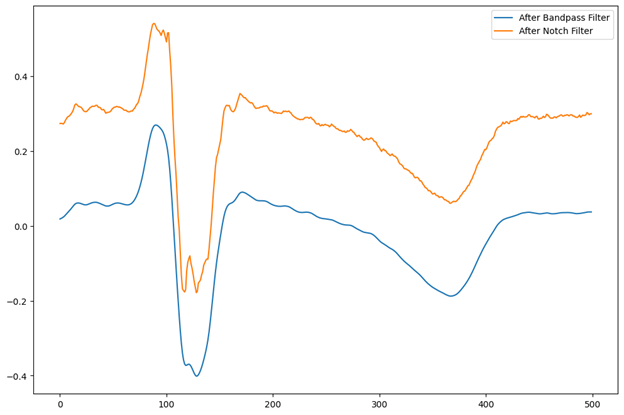}
        \caption{Second Layer}
        \label{fig:image2}
    \end{subfigure}%
    \begin{subfigure}[b]{0.33\textwidth}
        \includegraphics[width=\textwidth]{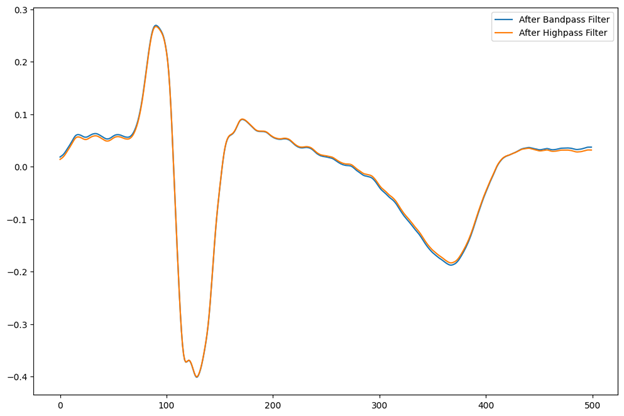}
        \caption{Third Layer}
        \label{fig:image3}
    \end{subfigure}
    \caption{Visionary V2 layer-by-layer Visualization}
    \label{fig:three_images}
\end{figure*}

The findings obtained after utilizing BioSPPy for signal processing to enhance ECG signal quality and create a format suitable for machine learning models did not satisfy the necessary accuracy criteria expected in medical data analysis. In search of a better ECG filter, we developed a multi-step filtration strategy called VisionaryV2 that performed better than previously used filtration techniques. ECG is a discrete time signal that is produced through routine sampling at regular intervals. For the study and design of discrete temporal systems that operate on such signals, including ECG data, a wide range of mathematical techniques are available \citep{9}.

Initial normalization involves reducing mechanical and hardware errors, particularly the powerline frequency noise that is common in ECG signals, in the pursuit of improved accuracy. It is imperative to address this noise because it introduces a significant, albeit uncommon, frequency into the ECG data. The raw signals are filtered to remove the desired frequency, such as the powerline frequency (e.g., 50Hz in the PTBDB dataset for Europe and many Asian countries, adjustable to 60Hz based on geographic location), using the first layer of Visionary V2, acting as a notch-filter.

The frequency ratio, which is obtained by dividing the particular frequency to be removed by the Nyquist frequency (in this study, 500Hz), serves as the basis for the notch-filter design. This frequency ratio and the IIR (Infinite Impulse Response) algorithm are used in the filter construction while a defined quality factor shapes the filter's bandwidth. The Nyquist frequency, a critical parameter in signal processing, represents the highest frequency accurately measurable by equipment at a specified sampling rate. Typically, it is half of the given sample rate, converting continuous data into a discrete sequence \citep{5}.

The mathematical foundation of IIR filters lies in their transfer function, H(s) which encapsulates the filter's behavior and is formulated as the ratio of two polynomials—namely, the numerator and the denominator. Mathematically, this is expressed as:

\begin{equation}
H(s) = \frac{{b_0 + b_1 s + b_2 s^2 + \ldots + b_M s^M}}{{a_0 + a_1 s + a_2 s^2 + \ldots + a_N s^N}}\label{eq}
\end{equation}

Here,  $b_i$ and $a_i$ denote the coefficients of the numerator and the denominator, respectively. The orders 
M and N of these polynomials define the filter's complexity and behavior. Due to their computational efficiency, IIR filters were preferred to FIR filters because they needed fewer coefficients to produce an equivalent FIR. The transformation of raw signals is made easier by the transfer function, which produces coefficient arrays. IIR filters are effectively implemented in the time domain with a simple difference equation for digital computation \citep{7,8}.

The standard form of the difference equation is:

\begin{align}
    y[n] &= b_0 x[n] + b_1 x[n-1] + \ldots + b_M x[n-M] \nonumber \\
         &- a_1 y[n-1] - \ldots - a_N y[n-N]
    \label{eq}
\end{align}

Here, $x[n]$ is the current input sample and $y[n]$ is the output sample where the hardware error is discarded. The converted signal is then passed through a zero-phase function to prevent phase-shifting. In biomedical signal processing, such as the ECG, this function preserves the temporal alignment of features in the filtered waveform with the unfiltered signal. It maintains critical timing information while improving accuracy and reducing artifacts.

A second layer of filtering, acting as a bandpass filter, is applied to the signal after initial processing. By allowing signals within a certain frequency range while rejecting others, this filter helps to smooth the signal and reduce overall noise in the signal. The Nyquist frequency is used to set the lowcut and highcut parameters, which are then optimized through empirical adjustments to 0.001 and 0.1 respectively for our dataset. Since the Laplace domain transfer function for a bandpass filter is a specific type of IIR transfer function, they have the same equation featuring different coefficients. The arrays produced by this stage are then further processed using a Butterworth filter, which was selected for its maximally flat frequency response possible within the passband and lowest possible amplitude ripple. From Fig. 3(b), it is clearly observed that sudden ups and downs in the notched signal are greatly mitigated and we get a smooth signal. To maintain accurate timing information, a zero-phase filter is then used. The bandpass filter also has a tendency to align the signal with the x-axis, resulting in an ECG signal that has been filtered and has a stable baseline, addressing amplitude variability brought on by various physiological conditions such as HBP, LBP, Electrolyte Level and age.

The final layer then receives the filtered signal and employs a highpass filter. A highpass filter is essential to address residual unwanted elements in the lower frequency range, such as baseline wander from patient movement and respiration, as well as environmental factors, despite the earlier application of a bandpass filter. By optimizing the frequency response and adapting dynamically, this highpass filter successfully gets rid of these issues. Its function is to complement the earlier stages by adding a second layer of filtering, enhancing the denoising of the original signal.

Then, using the R-peaks, mean, standard deviation, minimum, maximum, skew, and kurtosis of each heartbeat, singular normalized heartbeats are created. Finally, normalized heartbeats are scaled proportionally so that machine learning models can quickly identify the features needed for MI detection.

\subsection{Generated Image Dataset}

Since a suitable dataset for our image recognition model was not readily available, we generated our own from the existing PTBDB dataset. We first selected the raw numerical ECG data and applied the Visionary V2 filter for signal normalization and smoothing. This filter helped establish a baseline for ECG signals irrespective of age or gender, aiding in label prediction. We then segmented the filtered data into four-second intervals and converted these into initial images.

Subsequently, these images were transformed into grayscale and subjected to Sobel operator for edge detection. The Sobel outputs were made 8-bit and superimposed onto the original images in a 9:1 ratio to improve quality. These enhanced images were resized to $256 \times 256$ pixels, since the detection of patterns is of more importance in comparison to pixel-wise detail in ECG. Next, we look to increase the volume of our dataset using data augmentation methods to make our model more general to different kinds of ECG signals which follow the same image generation techniques as ours.

\begin{figure}[htbp]
\centerline{\includegraphics[scale=0.25]{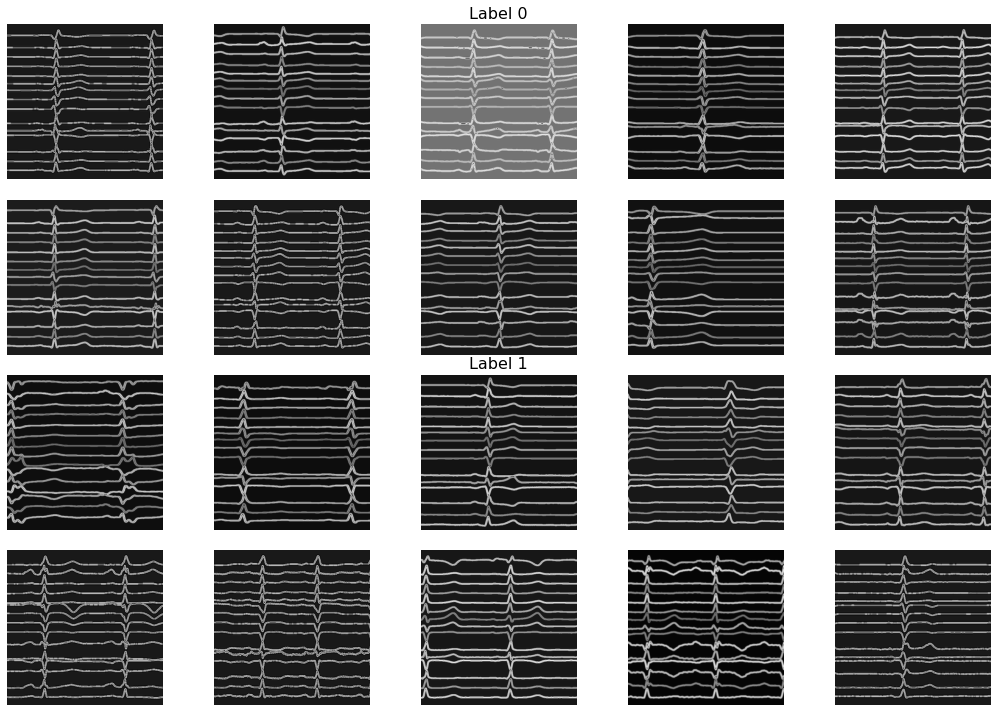}}
\caption{Generated Image Dataset}
\label{fig}
\end{figure}

Since ECG signals are sensitive medical reports that cannot be augmented randomly due to the risk of changing the actual signal orientation and pattern, we only chose the augmentation methods that did not distort the ECG shape, such as zooming, cropping, Gaussian blurring, and brightness and contrast adjustments and many more. This dataset is then split into train, test and validation sets. The validation set is created taking 15\% images from each of our classes. The total number of training images for our work was 42955 belonging to both classes as presented in Fig. 4. Since our training set consisted of a large amount of images, we had to use ImageDataGenerator to load our training set in batches.

\subsection{Image-based DL Models}

Custom CNN Model: Convolutional Neural Networks (CNNs) are a subset of artificial neural networks that are optimised for visual data processing. 

\begin{figure}[htbp]
\centerline{\includegraphics[width=0.4\textwidth, height=0.2\textheight]{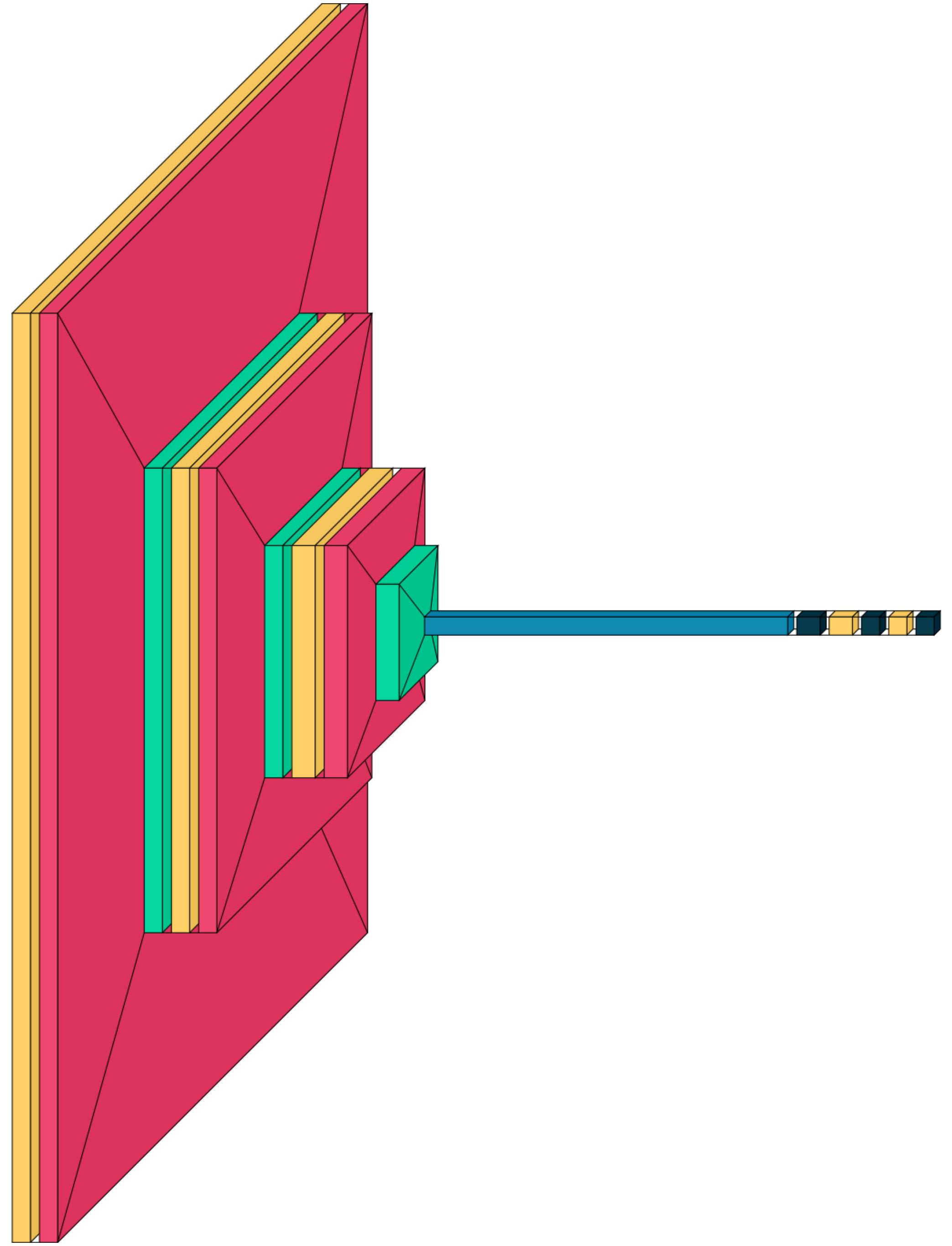}}
\label{fig}
\end{figure}
\begin{figure}[htbp]

\vspace{-5mm}

\centerline{\includegraphics[scale=0.1]{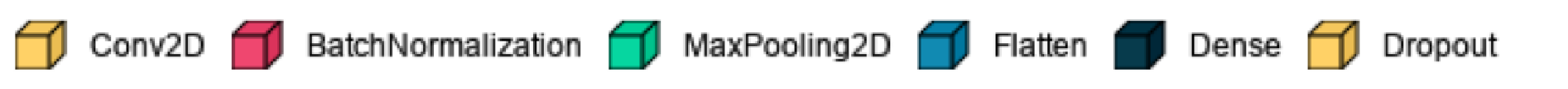}}
\caption{Custom CNN Architecture Visualization}
\label{fig}
\end{figure}

\vspace{-2mm}

In our work, we created a custom sequential model following the structure and architecture of a CNN model. The model takes in a $256 \times 256$ grayscale image as input in batches of 32 images and applies 32 filters on the input in the first convolutional block which keeps the same padding and uses the ReLu activation function. Following this, batch normalization is performed to stabilize the activations and accelerate the training process. Next, max pooling is applied, reducing the image to half its original size. There are 3 total convolutional blocks in our model ultimately resulting in a $32 \times 32$ image which is flattened into a 1D array before being input into the initial dense layer of 256 neurons with ReLu activation, followed by a dropout of 0.5. Next is another ReLu activated dense layer of 128 neurons and 0.5 dropout until finally we reach the output layer with sigmoid activation, which provides us with the prediction of the model. Our model uses the Adam optimizer with an initial learning rate of 0.001 and the binary crossentropy loss function. There were a total of 67,475,457 trainable parameters for the model. We also used ReduceLROnPlateau to update the learning rate on validation accuracy loss during model training.

Custom ResNet-50: ResNet-50 is comprised of 16 residual blocks that overall contains 48 convolutional layers, one MaxPool layer, and one average pool layer in its 50-layer design. The architecture includes skip connections, which simplifies gradient descent-related problems and facilitates training. 

\begin{figure}[htbp]
\centerline{\includegraphics[scale=0.037]{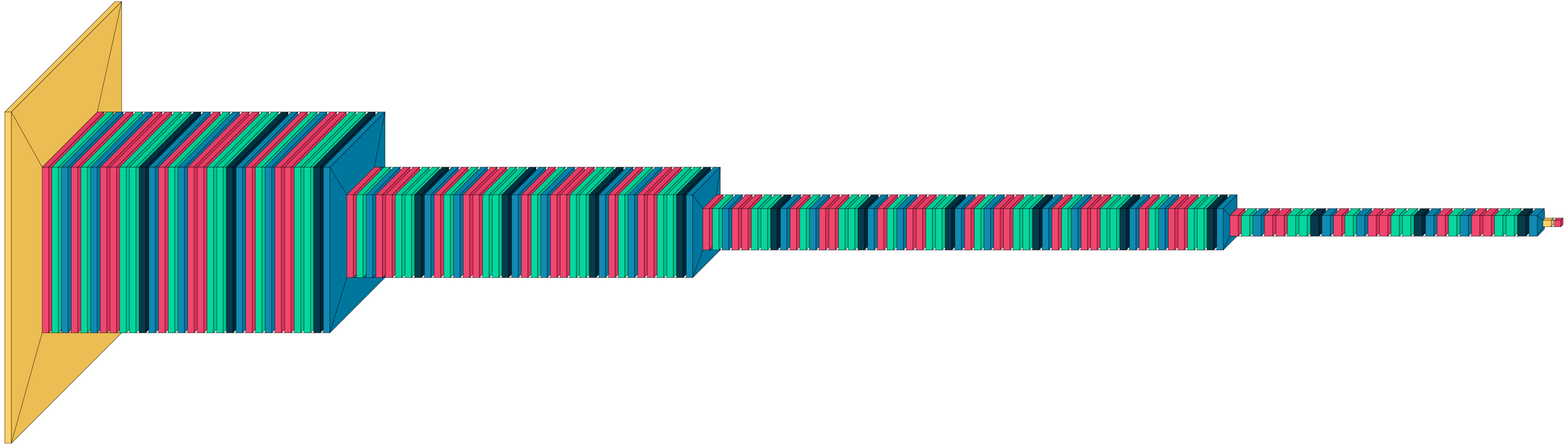}}
\label{fig}
\end{figure}

\vspace{-5mm}

\begin{figure}[htbp]
\centerline{\includegraphics[scale=0.15]{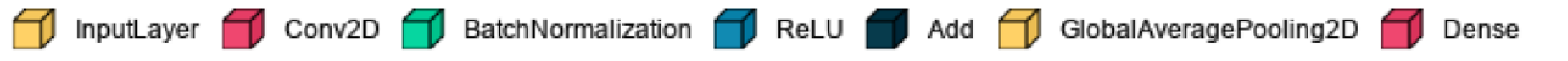}}
\caption{Custom ResNet-50 Architecture Visualization}
\label{fig}
\end{figure}

\vspace{-2mm}

In our custom implementation of the ResNet-50 architecture, we utilize a series of key components such as residual layers, convolutional layers, batch normalization, and pooling layers to construct the model. The input layer first takes in an image of $256 \times 256$ size and passes that to the initial convolution layer that applies 32 filters of $3 \times 3$ kernel size on the image before performing batch normalization on the resulting image which improves convergence speed of the ResNet model and subsequently ReLU activation function is used to help counter the vanishing gradient problem. The architecture incorporates residual blocks to facilitate the learning of residual mappings between inputs and outputs. The residual blocks consist of an initial convolution layer, batch normalization, a ReLU activation function followed by a shortcut connection. Post-shortcut, there exists another conv layer, a batch normalization with shortcut and a final batch normalization to sum up the residual block. After 16 residual blocks of similar design with varying number of filters, the Global Average Pooling brings the image dimensions down to 32. This leads to our output layer with the sigmoid activation function to provide the prediction from our model. The optimizer used for the model is Adam and the loss function is the binary crossentropy. Total trainable parameters were 5,560,545 here.

\subsection{XAI Integration}

The eXplainable Artificial Intelligence (XAI) subdiscipline was developed as a result of the increasing importance placed on the interpretability of complex models in the fast evolving fields of artificial intelligence (AI) and machine learning (ML). XAI seeks to provide machine learning models the ability to reveal insights into their decision-making processes in order to promote trust and enhance human comprehension.

\subsubsection{LIME}
LIME (Local Interpretable Model-agnostic Explanations) provides a framework for comprehending the forecasts of sophisticated ML models, which are frequently referred to as ``black-box'' systems. After being trained to anticipate how model prediction changes due to certain feature modifications, the explanation model ultimately offers a comprehensible illustration of how the original model generates its predictions \citep{12}.

\subsubsection{Grad-CAM}

Grad-CAM, or Gradient-weighted Class Activation Mapping, is a technique for illuminating the inner workings of DL models, particularly CNNs. Grad-CAM computes a rough localization map by making use of the gradients that have been propagated all the way back to the last convolutional layer of a particular model. The regions in the input data that are most important for the model's predictions are highlighted on this map \citep{13}.

\subsubsection{Integrated Gradients}

Integrated Gradients is a technique for deciphering the inner workings of DL systems, especially in figuring out how each input characteristic affects the prediction of the model. The method integrates gradients along a path that starts at a baseline (typically a neutral or ``zero'' input) and moves towards the actual input in question. Integrated Gradients quantify the prediction by quantitatively attributing each pixel in the image or each feature in the input to the model's output along this path by calculating the integral of the gradients \citep{14}.

\section{Evaluation Metrics and Findings}

\subsection{Numerical Data Results}

\begin{table}[H]
\caption{Base ML Models Performance Metrics}
\centering
\scriptsize 
\begin{tabular}{|c|c|c|c|c|}
\hline
\textbf{Model} & \textbf{{Type}}& \textbf{{Biosppy}}& \textbf{{EHC}}& \textbf{{VisionaryV2}} \\
\hline
\multirow{4}{*}{SVM} & Accuracy & 0.811 & 0.751 & 0.811 \\
                         & Precision & 0.811 & 0.819 & 0.811 \\
                         & Recall & 0.811 & 0.751 & 0.811 \\
                         & F1 Score & 0.726 & 0.763 & 0.726 \\
\hline
\multirow{4}{*}{KNN} & Accuracy & 0.811 & 0.942 & 0.855 \\
                         & Precision & 0.764 & 0.943 & 0.847 \\
                         & Recall & 0.811 & 0.942 & 0.855 \\
                         & F1 Score & 0.760 & 0.942 & 0.850 \\
\hline
\multirow{4}{*}{MLP} & Accuracy & 0.833 & 0.865 & 0.788 \\
                         & Precision & 0.820 & 0.862 & 0.654 \\
                         & Recall & 0.833 & 0.865 & 0.788 \\
                         & F1 Score & 0.788 & 0.861 & 0.715 \\
\hline
\multirow{4}{*}{Logistic Regression} & Accuracy & 0.822 & 0.778 & 0.888 \\
                         & Precision & 0.852 & 0.858 & 0.957 \\
                         & Recall & 0.855 & 0.860 & 0.955 \\
                         & F1 Score & 0.826 & 0.858 & 0.953 \\
\hline
\multirow{4}{*}{Decision Tree} & Accuracy & 0.711 & 0.918 & 0.933 \\
                          & Precision & 0.723 & 0.920 & 0.937 \\
                          & Recall & 0.711 & 0.918 & 0.933 \\
                          & F1 Score & 0.717 & 0.919 & 0.934 \\
\hline
\end{tabular}
\label{tab1}
\end{table}

\begin{table}[H]
\caption{Ensemble ML Models Performance Metrics}
\vspace{-5mm}
\begin{center}
\scriptsize
\begin{tabular}{|c|c|c|c|c|}
\hline
\textbf{Model Name} & \textbf{\textit{Type}}& \textbf{\textit{Biosppy}}& \textbf{\textit{EHC}}& \textbf{\textit{VisionaryV2}} \\
\hline
\multirow{4}{*}{Random Forest} & Accuracy & 0.877 & 0.971 & 0.955 \\
                         & Precision & 0.877 & 0.971 & 0.948 \\
                         & Recall & 0.877 & 0.971 & 0.955 \\
                         & F1 Score & 0.859 & 0.971 & 0.953 \\
\hline
\multirow{4}{*}{XGBoost$^{*}$} & Accuracy & 0.855 & 0.985 & 0.955 \\
                         & Precision & -- & 0.985 & -- \\
                         & Recall & -- & 0.985 & -- \\
                         & F1 Score & -- & 0.985 & -- \\
\hline
\multirow{4}{*}{AdaBoost} & Accuracy & 0.888 & 0.860 & 0.944 \\
                         & Precision & 0.884 & 0.858 & 0.943 \\
                         & Recall & 0.888 & 0.860 & 0.944 \\
                         & F1 Score & 0.886 & 0.858 & 0.942 \\
\hline
\multirow{4}{*}{Bagging} & Accuracy & 0.877 & 0.959 & 0.966 \\
                         & Precision & 0.870 & 0.959 & 0.967 \\
                         & Recall & 0.877 & 0.959 & 0.966 \\
                         & F1 Score & 0.864 & 0.959 & 0.965 \\
\hline
\end{tabular}

\vspace{1mm}
{\scriptsize $^{*}$Precision, recall and F1 score of XGBoost on the BioSPPy and Visionary V2 features were not preserved from the original experiments and are omitted.}
\label{tab1}
\end{center}
\end{table}

\vspace{-5mm}

\subsection{Image Data Results}

\vspace{-3mm}

\begin{table}[H]
\caption{Image Performance Metrics}
\vspace{-5mm}
\begin{center}
\scriptsize
\begin{tabular}{|c|c|c|c|}
\hline
\textbf{Model Name} & \textbf{\textit{Result Type}}& \textbf{\textit{Healthy}}& \textbf{\textit{MI}} \\
\hline
\multirow{5}{*}{Custom CNN} & Precision & 0.987 & 0.986 \\
                    & Recall & 0.986 & 0.987 \\
                    & F1 Score & 0.986 & 0.987 \\
                    & Sensitivity & 0.986 & 0.987 \\
                    & Specificity & 0.987 & 0.987 \\
                         
\hline
\multirow{5}{*}{Custom ResNet-50} & Precision & 1.000 & 0.999 \\
                    & Recall & 0.999 & 1.000 \\
                    & F1 Score & 0.999 & 0.999 \\
                    & Sensitivity & 0.999 & 1.000 \\
                    & Specificity & 1.000 & 0.999 \\

\hline
\end{tabular}
\label{tab1}
\end{center}
\end{table}

\vspace{-5mm}

\subsection{Comparative Analysis}

Our objective was to attain optimal accuracy with a lightweight model. We evaluated five base models (SVM, KNN, MLP, Logistic Regression, Decision Tree) and found that Decision Tree yielded the highest accuracy 93.3\% and F1 score 93.4\% compared to other filtration methods. Subsequently, we pursued ensemble models using Decision Tree as the base, including Random Forest, XGBoost, AdaBoost, and Bagging. Among these, Bagging Classifier emerged as the top performer with an accuracy of 96.6\% and an F1 score of 96.5\%, followed by Random Forest with 95.5\% accuracy and 95.3\% F1 score.

In the case of image classification models, our custom ResNet-50 model performed better than the custom CNN model with 99.9\% precision, 100\% recall and 99.9\% F1 score compared to the 98.6\% precision, 98.7\% recall and 98.7\% F1 score of the latter model. From the confusion matrix (Fig. 7), it can be seen that the  ResNet-50 model mispredicted only 1 image out of 7000 whereas CNN model mispredicted 91 out of 7000 images.

\begin{figure}[h] 
  \centering 

  \begin{minipage}{0.49\linewidth}
    \includegraphics[width=\linewidth]{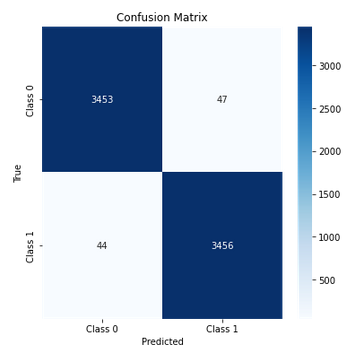}
    \label{fig:image1}
  \end{minipage}
  \hfill 
  \begin{minipage}{0.49\linewidth}
    \includegraphics[width=\linewidth]{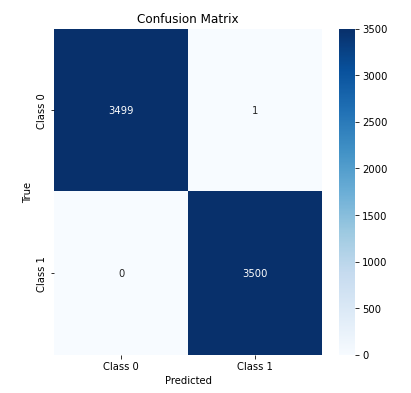}
    
    \label{fig:image2}
  \end{minipage}
  \hfill
\caption{Confusion Matrices}
\end{figure}

\vspace{-5mm}

\section{Insights and Interpretations}

XAI is implemented to identify and comprehend the reasoning behind the results and outputs generated by Machine Learning Algorithms to make them interpretable by humans. We have implemented the LIME, Grad-CAM and Integrated Gradients techniques to verify if our models selected the proper features from the input images. LIME provided a general mask by using a local surrogate model, illuminating regions of the input image that significantly impacted the model's decision-making. The Grad-CAM technique visualized important areas in the input images as a heatmap generated by stacking marked layers, highlighting feature relevance. Integrated Gradients offered a granular understanding, assessing the model's focus on correct features for MI diagnosis on a pixel-wise basis. Our XAI results suggest that the models attend to the ECG waveform itself, since only the signal regions were highlighted while the empty background was ignored, indicating appropriate feature selection. Although we are unable to classify among the various classes of MI due to data scarcity, we aim to overcome this limitation by generalizing the outcome of the XAI techniques by finding a common pattern from each individual lead. 

\section{Challenges and Future Works}

The results generated from our study provide a promising foundation for the use of machine learning and deep learning models in the identification of cardiovascular diseases such as MI. But there are a few possible directions for future study that could help us improve and build on our work.

To improve performance and take advantage of potential advantages, we intend to investigate alternative deep learning and machine learning models. We also plan to create custom models that are specialized in learning ECG features for better classification.

We plan to improve model accuracy by including additional factors like patient demographics, medical history, and specific ECG features such as QRS complex and ST segment elevation. While data scarcity is a concern, these enhancements could not only improve predictions but also help identify specific types of MI based on abnormal ECG leads.

Finally, the integration of these models into a medical environment will be a significant next step. It would be essential to evaluate the system's usefulness and efficacy in an actual clinical setting as a secondary evaluator.

With these prospective endeavors, we expect to continue advancing the field of machine and deep learning in cardiology, with the ultimate objective of enhancing MI detection and patient outcomes.

\section{Conclusion}

In conclusion, this study explores the potential of machine learning and deep learning techniques in the diagnosis of myocardial infarction using ECG signals. The research involved collecting data from PTBDB dataset, preprocessing and extracting features from the signals, and developing and evaluating multiple models on this data. The results indicate that certain models like Random Forest and Bagging achieve high accuracy in detecting MI from the ECG signals. 

The research was extended by generating an image dataset from the ECG signals and building custom models like CNN and ResNet-50 for image classification. The deep learning models demonstrated excellent performance in pattern detection, with over \(99\%\) accuracy in detecting MI. Techniques like Grad-CAM and LIME were also implemented to visualize the predictions of the deep learning models.

 With further research to validate and improve the models on larger diverse datasets, as well as integration into real clinical environments, these AI techniques could enable earlier diagnosis of MI and help reduce associated mortality rates. The models' ability to provide personalized risk assessments could also assist in preventative screening and treatment. In summary, this research serves as a promising step towards developing AI systems that can augment clinicians' capabilities in accurately diagnosing cardiovascular diseases.

\bibliographystyle{model5-names}
\bibliography{refs}

\end{document}